\documentclass[%
 reprint,
superscriptaddress,
 amsmath,amssymb,
 aps,
]{revtex4-2}

\usepackage{graphicx}% Include figure files
\usepackage{dcolumn}% Align table columns on decimal point
\usepackage{bm}% bold math
\usepackage{verbatim}
\usepackage{xcolor}

\newcommand{\minus}{\scalebox{0.75}[1.0]{$-$}}

\begin{document}

\preprint{APS/123-QED}

\title{Completely characterizing multimode nonlinear-optical quantum processes}% Force line breaks with \\
\author{Geunhee Gwak}
\affiliation{Department of Physics, Korea Advanced Institute of Science and Technology (KAIST), Daejeon 34141, Korea}

\author{Chan Roh}
\affiliation{Department of Physics, Korea Advanced Institute of Science and Technology (KAIST), Daejeon 34141, Korea}

\author{Young-Do Yoon}
\affiliation{Department of Physics, Korea Advanced Institute of Science and Technology (KAIST), Daejeon 34141, Korea}

\author{M.S. Kim}
\affiliation{Blackett Laboratory, Imperial College London, London SW7 2AZ, United Kingdom}

\author{Young-Sik Ra}
\email{youngsikra@gmail.com}
\affiliation{Department of Physics, Korea Advanced Institute of Science and Technology (KAIST), Daejeon 34141, Korea}

\date{\today}% It is always \today, today,

\begin{abstract}
Complete characterization of a multimode optical process has paved the way for understanding complex optical phenomena, leading to the development of novel optical technologies. Until now, however, characterizations have mainly focused on a linear-optical process, despite the plethora of multimode nonlinear-optical processes crucial for photonic technologies. Here we report the experimental characterization of multimode nonlinear-optical quantum processes by obtaining the full information needed to describe them while satisfying the necessary physical condition. Specifically, to characterize a second-order nonlinear-optical process of parametric downconversion, we determine the amplification and noise matrices of multimode field quadratures. The full information allows us to factorize the multimode process, leading to the identification of eigenquadratures and their associated amplification and noise properties. Moreover, we demonstrate the broad applicability of our method by characterizing various nonlinear-optical quantum processes, including cluster state generation, mode-dependent loss with nonlinear interaction, and a quantum noise channel. Our method, by providing a versatile technique for characterizing a nonlinear-optical process, will be beneficial for developing scalable photonic technologies.
\end{abstract}

\maketitle

%\tableofcontents

%\section{\label{sec:level1}Introduction}

Characterizing an unknown physical process is a primary endeavor of science, deemed crucial not only for attaining the correct understanding but also for facilitating precise predictions. In optics, the pertinent task is to determine the interactions of multiple optical modes, known as the input-output relation \cite{popoffTmat,boydbook,weedbrook,fabre2020modes,moon2023measuring}. Extensive studies on the multimode characterization have been conducted in both classical and quantum optics fields. For classical optics, characterization based on a transmission (reflection) matrix has enabled a systematic analysis of complex light scattering \cite{popoffTmat,Shapingcao}, advancing optical techniques for imaging \cite{bertolotti2022imaging,mosk2012controlling,kimmaxenergy,Baek2023}, sensing \cite{Fisherinfo,Pai:2021aa} and computing \cite{Computinggigan}. For quantum optics, characterization of multiport optical circuits, whether in bulk \cite{quantumcomputing} or integrated \cite{universallinear} optics, or even within scattering media \cite{Goel:2024aa}, has allowed the engineering of quantum interference for advancing quantum technologies \cite{complexlib,Integratedwang}. However, this method is intrinsically limited to characterizing linear-optical processes.

On the other hand, a nonlinear-optical quantum process such as parametric downconversion \cite{PDC,boydbook,30years} plays a crucial role in photonic technologies. It facilitates parametric amplification \cite{alloptical}, phase conjugation \cite{Pereira:1994aa,phaseconjugation}, and generation of quantum light (entangled photons \cite{PDC} and squeezed light \cite{30years}). Recent experimental progresses have extended the nonlinear process into the multimode regime \cite{wright2022physics}, paving the way for scalable quantum technologies. A highly multimode nonlinear process in bulk \cite{cainatcom,3dcluster,ProgramupconversionFederico,Barakat:2024aa} or integrated \cite{Fewcylesqueezingnehra,integratedgraph} optics offers scalability for quantum computing \cite{alloptical}, quantum communication \cite{Kovalenko:2021ej,Liu:2022eu,Notarnicola:2024aa}, and quantum sensing \cite{Lemos:2014aa,wide11interferometer}. In accordance with these recent advances, the complete characterization of a multimode nonlinear-optical process becomes increasingly indispensable.

Interestingly, this nonlinear-optical process exhibits much richer features than the linear one, necessitating a generalized methodology for the complete characterization \cite{weedbrook, Lvovsky:2009fr, Wang:2013kl, Fiurasek:2015iq, Teo:2021ks}. In contrast to the linear process, where the complex amplitudes of electric fields are sufficient in determining the output field, both the complex amplitudes and their conjugations (equivalently, multimode field quadratures~\cite{fabre2020modes}) contribute to the output field: this renders the conventional linear-optical characterization method insufficient. Moreover, the nonlinear process intrinsically introduces noise (either quantum or classical) to the output field. Lastly, the characterization result should obey the physical condition governed by quantum mechanical principles~\cite{weedbrook}. However, the complete experimental characterization of a multimode nonlinear-optical quantum process still remains elusive \cite{SET2,Ra:2017ia,Huo:2020il,JSAphotonpair}.

In this article, we experimentally characterize a multimode nonlinear-optical quantum process by obtaining the complete information. More specifically, we determine an amplification matrix (revealing the interplay among multimode field quadratures) and a noise matrix (identifying noise induced by environmental interactions), while satisfying the necessary physical condition. Crucially, the noise matrix provides a general framework to characterize non-unitary evolutions in realistic experiments (e.g., mode mismatch~\cite{Roh:2021bv} and optical loss~\cite{modeloss}). By analyzing the characterization results, we successfully identify eigenquadratures and their associated amplification and noise properties. Furthermore, we showcase broad applicability of our method by characterizing nonlinear-optical quantum processes including cluster state generation, mode-dependent loss with nonlinear interaction, and a quantum noise channel.

We start by establishing the input-output relationship for electric field amplitudes. In classical optics, the input-output relation through linear optics is given by a transmission matrix \textbf{\textit{T}} \cite{popoffTmat}. For an input field amplitude $E_m$ in an optical mode $m$, the transmission matrix determines the output field amplitude $E'_n$ by $\sum_m T_{nm} E_m$. On the other hand, a nonlinear-optical process, such as parametric downconversion, involves both the field amplitude $E_m$ and the conjugate amplitude $E^*_m$ to determine the output field amplitude \cite{boydbook}: $E'_n=\sum_m (G_{nm} E_m + H_{nm} E^*_m)$. Here \textbf{\textit{G}} and \textbf{\textit{H}} are complex-valued matrices, generally independent of each other, and the new term \textbf{\textit{H}} offers nonlinear-optical effects such as parametric amplification \cite{alloptical} and phase conjugation \cite{Pereira:1994aa,phaseconjugation}.

\begin{figure}
\includegraphics[width=90mm,scale=0.7]{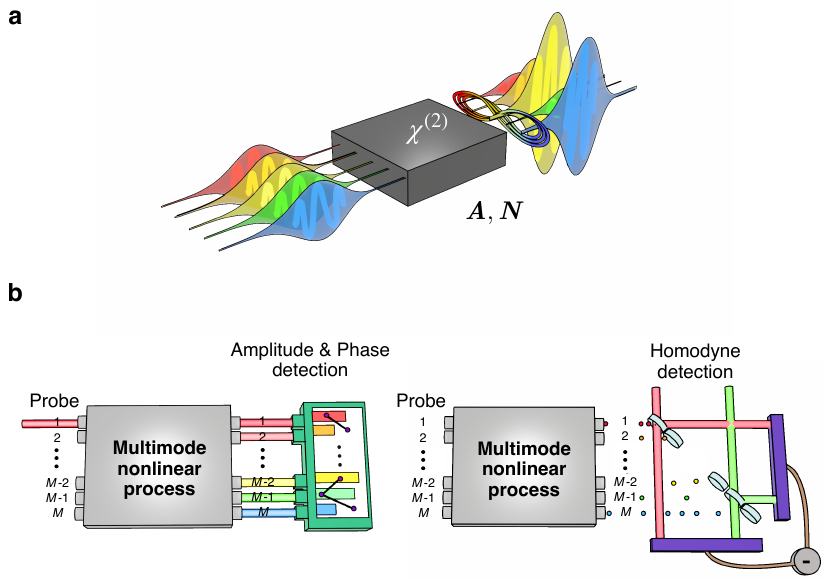}% Here is how to import EPS art
\caption{\textbf{A multimode nonlinear-optical quantum process.} \textbf{a,} Unlike a linear-optical process, a second-order nonlinear-optical process ($\chi^{(2)}$) induces optical amplification and noise in multimode fields. This nonlinear process is fully characterized by the amplification matrix \textbf{\textit{A}} and the noise matrix \textbf{\textit{N}} in Eq.~(\ref{eq1}). \textbf{b,} Our method for completely characterizing a multimode nonlinear-optical quantum process. We first inject coherent states in sequence as a probe into an unknown multimode nonlinear process, and measure the output mean quadrature vector (left). Next, we use the multimode vacuum state as the probe, and measure the output state by using homodyne detection (right). The former provides information about \textbf{\textit{A}}, and the latter provides additional information about \textbf{\textit{N}}, enabling the complete characterization of the nonlinear-optical process and thereby facilitating the prediction of the evolution of general multimode quantum states.}
\label{fig1}
\end{figure}

To fully account for the quantum characteristics of light~\cite{fabre2020modes}, we employ photon annihilation ($\hat{a}_m$) and creation ($\hat{a}^\dagger_m$) operators which generalize the field amplitude $E_m$ and the conjugate amplitude $E_m^*$ in classical optics. Multimode electric fields then can be described by using a vector of quadrature operators $\hat{\textbf{\textit{q}}}=(\hat{x}_1, \hat{x}_2, ...\hat{x}_M, \hat{p}_1,\hat{p}_2,...,\hat{p}_M)^T$, where $\hat{x}_m = \hat{a}_m + \hat{a}^\dagger_m$ and $\hat{p}_m = (\hat{a}_m - \hat{a}^\dagger_m)/i$ with the commutation relation of $[\hat{x}_m,\hat{p}_m]=2i$. The full characteristics of the nonlinear-optical process finally appear in the input-output relations of a mean quadrature vector $\textbf{\textit{q}}$ ($q_m = \langle \hat{q}_m \rangle$, corresponding to multimode field quadratures) and a covariance matrix \textbf{\textit{V}} ($V_{mn} = \langle \{\Delta \hat{q}_m, \Delta \hat{q}_n\}/2 \rangle$, describing noise properties); this is because the second-order nonlinear-optical process is a Gaussian quantum process requiring information up to the second-order moments \cite{weedbrook}. The input-output relations are
\begin{align}
\textbf{\textit{q}}^\prime&=~\textbf{\textit{A}}\textbf{\textit{q}}\ \ \ \ \ \ \   \nonumber\\
\textbf{\textit{V}}^\prime&=~\textbf{\textit{AVA}}^T + \textbf{\textit{N}},
\label{eq1}
\end{align}
which are determined by an amplification matrix \textbf{\textit{A}} (a real $2M \times 2M$ matrix) and a noise matrix \textbf{\textit{N}} (a real symmetric $2M \times 2M$ matrix), see Fig.~\ref{fig1}a. By providing the complete information about both linear and nonlinear optical processes, \textbf{\textit{A}} reveals the interplay among multimode field quadratures, while \textbf{\textit{N}} illustrates the noise introduced by environmental interactions. \textbf{\textit{A}} is related to the aforementioned \textbf{\textit{G}} and \textbf{\textit{H}} by $\textbf{\textit{A}} = \begin{pmatrix} \textrm{Re}[\textbf{\textit{G}}+\textbf{\textit{H}}] & \textrm{Im}[\textbf{\textit{H}}-\textbf{\textit{G}}] \\ \textrm{Im}[\textbf{\textit{G}}+\textbf{\textit{H}}] & \textrm{Re}[\textbf{\textit{G}}-\textbf{\textit{H}}]
\end{pmatrix}$. Importantly, \textbf{\textit{A}} and \textbf{\textit{N}} should satisfy the physical condition originating from the Heisenberg uncertainty relation, which is translated to \cite{weedbrook}

\begin{align}
\textbf{\textit{N}}+i\boldsymbol{\mathit{\Omega}}-i\textbf{\textit{A}}\boldsymbol{\mathit{\Omega}} \textbf{\textit{A}}^T \succeq 0,
\label{eq2}
\end{align}
where $\boldsymbol{\mathit{\Omega}} = \begin{pmatrix} \textbf{\textit{0}}_M & \textbf{\textit{I}}_M \\ -\textbf{\textit{I}}_M & \textbf{\textit{0}}_M
\end{pmatrix}$ consisting of the zero matrix $\textbf{\textit{0}}_M$ and the identity matrix $\textbf{\textit{I}}_M$ of dimension $M$.
Note that Eq.~(\ref{eq2}) represents a more stringent condition than the physical condition for the output covariance matrix when the input is the multimode vacuum state ($\textbf{\textit{AA}}^T + \textbf{\textit{N}} + i\boldsymbol{\mathit{\Omega}} \succeq 0$)~\cite{weedbrook}, i.e., $\textbf{\textit{AA}}^T+ \textbf{\textit{N}} + i\boldsymbol{\mathit{\Omega}} \succeq 
\textbf{\textit{N}}+i\boldsymbol{\mathit{\Omega}}-i\textbf{\textit{A}}\boldsymbol{\mathit{\Omega}} \textbf{\textit{A}}^T \succeq 0$.

In quantum information theory, this multimode nonlinear-optical process is classified as a Bosonic Gaussian channel \cite{weedbrook, caruso2008multi}. A general form of a Bosonic Gaussian channel  includes an additional mean quadrature vector ($\textbf{\textit{d}}$) in the first line of Eq.~(\ref{eq1}): $\textbf{\textit{q}}^\prime=~\textbf{\textit{A}}\textbf{\textit{q}} + \textbf{\textit{d}}$. Note that $\textbf{\textit{d}}$ can be easily determined by setting $\textbf{\textit{q}}=\textbf{\textit{0}}$, and in a multimode process without external mean fields, $\textbf{\textit{d}}=\textbf{\textit{0}}$ is naturally satisfied. The formalism based on $\textbf{\textit{A}}$, $\textbf{\textit{N}}$, and $\textbf{\textit{d}}$ can fully characterize both Gaussian unitary and non-unitary evolutions~\cite{weedbrook}, and the characterization result can predict the evolution of general multimode quantum states including both Gaussian and non-Gaussian quantum states~\cite{caruso2008multi,Ra:2020gg}.
%Additionally, the formalism based on $\textbf{\textit{A}}$ and $\textbf{\textit{N}}$ can characterize both Gaussian unitary and non-unitary evolutions~\cite{weedbrook}. Finally, the evolution of quadrature operators ($\hat{\textbf{\textit{q}}}$) can be determined through Gaussian dilations~\cite{caruso2008multi}, enabling the prediction for general input states~\cite{Ra:2020gg}.

\begin{figure*}[tbp]
\includegraphics[width=180mm,scale=0.5]{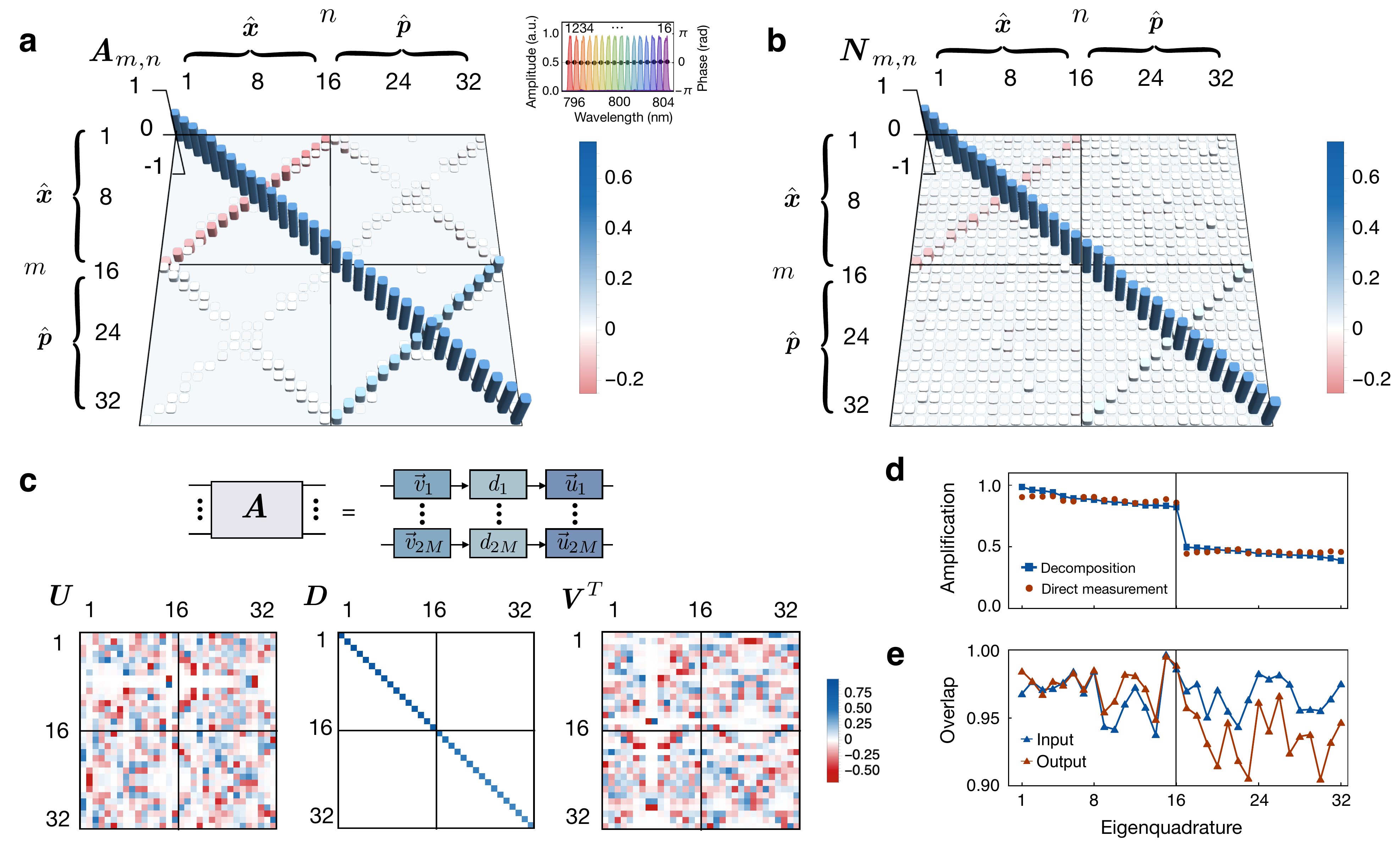}% Here is how to import EPS art
\caption{\textbf{Experimental characterization of a 16-mode nonlinear-optical quantum process.} \textbf{a,b,} The nonlinear process is completely characterized by obtaining ({\textbf{a}}) amplification matrix \textbf{\textit{A}} and (\textbf{b}) noise matrix  \textbf{\textit{N}}. $\hat{\textbf{\textit{x}}}=(\hat{x}_1, \hat{x}_2, ..., \hat{x}_{16})^T$, $\hat{\textbf{\textit{p}}}=(\hat{p}_1, \hat{p}_2, ..., \hat{p}_{16})^T$, and the inset shows the wavelength distribution of the 16 modes. \textbf{c,} A singular value decomposition of \textbf{\textit{A}} into $\textbf{\textit{U}}\textbf{\textit{D}}\textbf{\textit{V}}^T$. The decomposition allows us to understand the multimode process $\textbf{\textit{A}}$ as independent amplification processes (see the inset): the input (output) eigenquadratures $\{\vec{v}_m\}$ ($\{\vec{u}_m\}$) are the column vectors of $\textbf{\textit{V}}$ ($\textbf{\textit{U}}$), where the associated amplifications $\{d_m\}$ are the diagonals of $\textbf{\textit{D}}$.
The eigenquadratures are plotted in more detail in Extended Data Fig.~\ref{ext_fig1} and \ref{ext_fig2}. \textbf{d,} Amplifications of the eigenquadratures:  prediction from the decomposition result $\{d_m\}$ and direct experimental measurement by injecting probes in eigenquadratures. \textbf{e,} The overlap between experimentally generated eigenquadratures (input: $\{\vec{v}_m^{\textrm{(exp)}} \}$, output: $\{\vec{u}_m^{\textrm{(exp)}} \}$) and the eigenquadratures obtained by the decomposition, quantified by $( \vec{v}_m^{\textrm{(exp)}} \cdot \vec{v}_m )^2$ and $( \vec{u}_m^{\textrm{(exp)}} \cdot \vec{u}_m )^2$.
}\label{fig2}
\end{figure*}

Figure~\ref{fig1}b illustrates the schematic of obtaining the amplification matrix $\textbf{\textit{A}}$ and the noise matrix $\textbf{\textit{N}}$ of a multimode nonlinear-optical quantum process. First, we probe the process by using an input coherent state of a mean quadrature vector $\textbf{\textit{q}}$ and measure the output mean quadrature vector $\textbf{\textit{q}}^\prime$. To identify all matrix elements of $\textbf{\textit{A}}$, we use an input quadrature vector $\textbf{\textit{q}}^{(n)}$ in sequence ($n=1,\dots,2M$ and $\textit{q}_k^{(n)}=q \delta_{kn}$), and, with each output quadrature vector $\textbf{\textit{q}}^{\prime(n)}$, $\textit{A}_{mn}$ is determined by calculating $\textit{q}^{\prime{(n)}}_m/q$. Next, we probe the nonlinear process using the multimode vacuum state (exhibiting $q_m=0$ and $V_{mn}=\delta_{mn}$). The output state exhibits $\textbf{\textit{q}}^\prime=0$ and $\textbf{\textit{V}}^\prime=\textbf{\textit{A}}\textbf{\textit{A}}^T + \textbf{\textit{N}}$ from Eq.~(\ref{eq1}), which is measured by quadrature operators in different combinations ($\hat{x}_i$, $\hat{p}_i$, $\hat{x}_i+\hat{x}_j$, $\hat{p}_i+\hat{p}_j$, and $\hat{x}_i+\hat{p}_j$ for $i,j=1,\dots,M$). Based on the quadrature measurement outcomes, we finally reconstruct $\textbf{\textit{N}}$ satisfying the physical condition in Eq.~(\ref{eq2}) by applying the maximum likelihood estimation technique (Supplementary Information)~\cite{Lvovsky:2009fr}.

We employ this method to characterize multimode nonlinear-optical quantum processes by a synchronously pumped optical parametric oscillator (SPOPO)~\cite{cainatcom,3dcluster}. Detailed experimental setup is explained in Supplementary Information. SPOPO introduces a highly multimode nonlinear-optical process in the frequency domain, involving interactions of hundreds of optical modes \cite{patera} and generations of intricately entangled quantum states \cite{3dcluster}. Importantly, it can realize various multimode nonlinear interactions in a reconfigurable manner by choosing appropriate time-frequency mode bases~\cite{fabre2020modes,cainatcom,3dcluster}. We leverage such advantage to realize and characterize various types of multimode nonlinear-optical quantum processes.

\begin{figure*}
\includegraphics[width=180mm,scale=0.4]{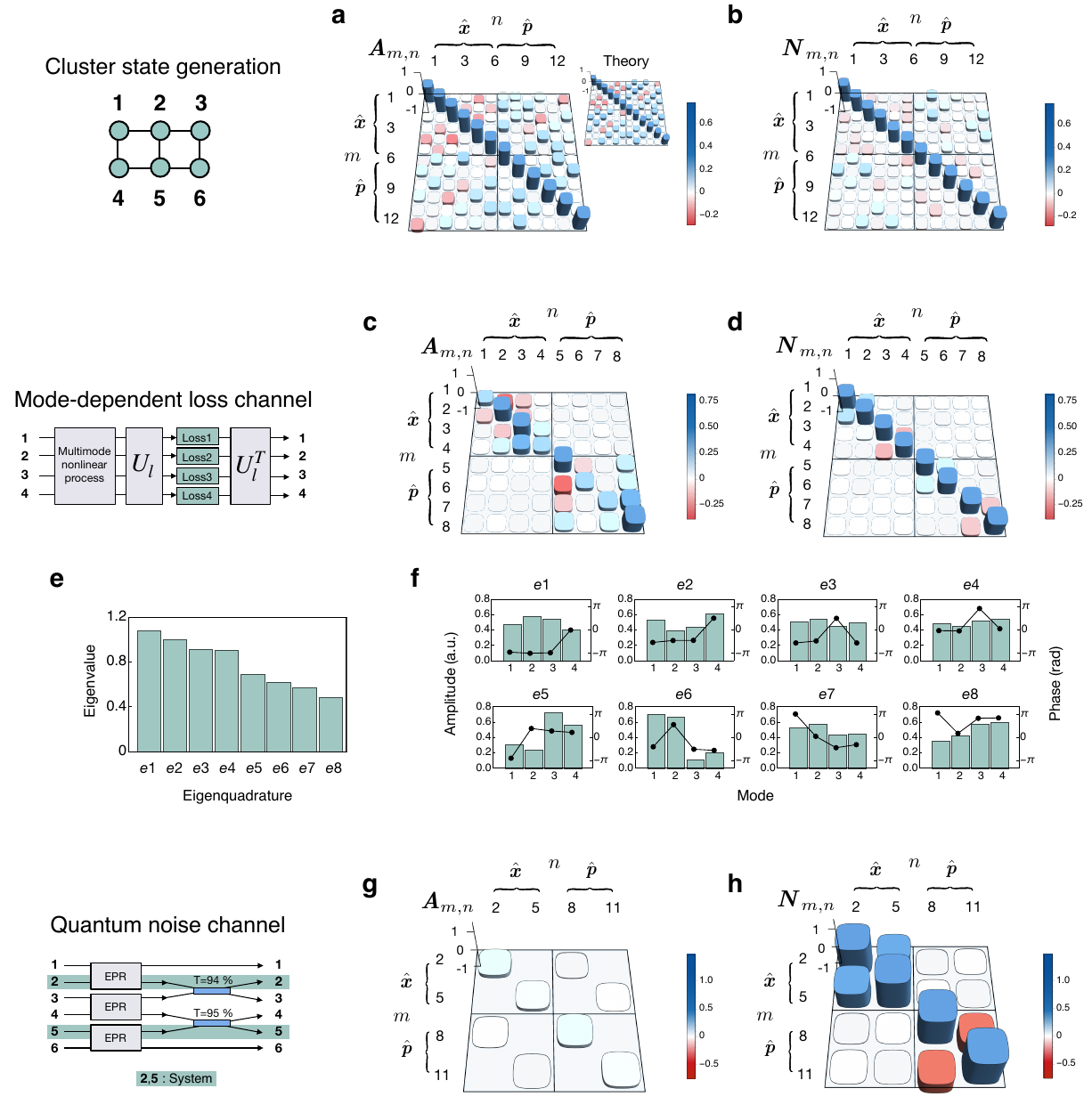}
\caption{\textbf{Experimental characterization of various multimode nonlinear-optical quantum processes.} \textbf{a,b,} Cluster state generation. \textbf{a} and \textbf{b} represent the experimentally obtained \textbf{\textit{A}} and \textbf{\textit{N}}, respectively. The inset in \textbf{a} shows the theoretical prediction for comparison. \textbf{c-f,} Mode-dependent loss with nonlinear interaction (\textbf{c}: \textbf{\textit{A}}, \textbf{d}: \textbf{\textit{N}}). \textbf{\textit{N}} exhibits correlated noise having non-vanishing off-diagonals. \textbf{e,f,} Eigendecomposition of \textbf{\textit{N}} (\textbf{e}: eigenvalues, \textbf{f}: eigenquadratures). In \textbf{f}, eigenquadratures are represented by complex superpositions of the four optical modes. 
\textbf{g,h,} Quantum noise channel (\textbf{g}: \textbf{\textit{A}}, \textbf{h}: \textbf{\textit{N}}). The off-diagonals in \textbf{\textit{N}} are sufficiently large to introduce quantum-correlated noise to the output. For the vacuum input, the output exhibits a negative partial-transposition eigenvalue of $\minus 0.37$~\cite{Duan:2000fw}, making the process an entanglement-generating channel. See Supplementary Information for detailed methods for realizing the multimode nonlinear-optical processes.
}
\label{fig3}
\end{figure*}

%{\label{sec:level2}Result}

We begin with investigating a multimode process in 16 frequency-band modes, which are equally spaced from 796 nm to 804 nm (the inset of Fig.~\ref{fig2}a). The complete information is represented by \textbf{\textit{A}} and \textbf{\textit{N}}, which are shown in Fig.~\ref{fig2}a,b. The amplification matrix \textbf{\textit{A}} reveals that the multimode process is a set of phase-insensitive amplifiers~\cite{boydbook,Pereira:1994aa}, which amplifies a coherent state at a given frequency mode (e.g., $m$) while generating a phase-conjugated coherent state at the conjugate mode ($17-m$). This nonlinear-optical process is also called difference frequency generation~\cite{boydbook}. The phase conjugation is highlighted by the non-zero off-diagonal elements showing the opposite signs between $\textbf{\textit{A}}_{ \hat{\textbf{\textit{x}}}, \hat{\textbf{\textit{x}}} }$ (the upper-left quadrant of $\textbf{\textit{A}}$) and $\textbf{\textit{A}}_{ \hat{\textbf{\textit{p}}}, \hat{\textbf{\textit{p}}} }$ (the lower-right quadrant of $\textbf{\textit{A}}$): the discrepancy between $\textbf{\textit{A}}_{ \hat{\textbf{\textit{x}}}, \hat{\textbf{\textit{x}}} }$ and $\textbf{\textit{A}}_{ \hat{\textbf{\textit{p}}}, \hat{\textbf{\textit{p}}} }$ is a unique signature of a nonlinear-optical process~\cite{weedbrook,boydbook}. The diagonals of \textbf{\textit{A}} do not exceed one due to the optical loss, which also causes a non-zero noise matrix \textbf{\textit{N}} (see Fig.~\ref{fig2}b). 
% (average of 42.5 \% including mode mismatch).
By substituting $\textbf{\textit{A}}$ and $\textbf{\textit{N}}$ into Eq.~(\ref{eq1}), one can predict the output covariance matrix for the input vacuum state: the results reveal the generation of eight pairs of Eienstein-Podolsky-Rogen (EPR) entanglement, exhibiting negative eigenvalues under partial transposition ($\minus0.27, \minus0.26, \minus0.24, \minus0.25, \minus0.26, \minus0.27, \minus0.25, \minus0.28$)~\cite{Duan:2000fw}.

The full acquisition of $\textbf{\textit{A}}$ allows us to analyze the multimode nonlinear-optical process through singular value decomposition (Fig.~\ref{fig2}c): $\textbf{\textit{A}} = \textbf{\textit{U}}\textbf{\textit{D}}\textbf{\textit{V}}^T$, where $\textbf{\textit{U}}$ and $\textbf{\textit{V}}$ are real orthogonal matrices, and $\textbf{\textit{D}}$ is a non-negative diagonal matrix. The column vectors of $\textbf{\textit{V}}$ and $\textbf{\textit{U}}$ represent the input ($\{\vec{v}_m\}$) and the output ($\{\vec{u}_m\}$) eigenquadratures, respectively, and the diagonals of $\textbf{\textit{D}}$ indicate the amplifications of the associated eigenquadratures. Figure~\ref{fig2}d shows the obtained amplification values which exhibit a clear gap between the first half (being amplified) and the second half (being deamplified) of the eigenquadratures. It indicates that the nonlinear-optical process functions as a set of phase-sensitive amplifiers for the eigenquadratures (Extended Data Fig.~\ref{ext_fig1})~\cite{boydbook,alloptical}. To experimentally confirm the validity of this decomposition, we directly inject a probe in each eigenquadrature and measure the corresponding amplification and output quadrature. The direct measurement results, shown in Fig.~\ref{fig2}d,e, agree well with the decomposition results.

The characterization method is broadly applicable to various multimode nonlinear-optical quantum processes. Detailed experimental methods for their implemention are described in Supplementary Information. In Fig.~\ref{fig3}a,b, we investigate a multimode nonlinear-optical process for generating a cluster state~\cite{3dcluster}. The amplification matrix $\textbf{\textit{A}}$ agrees well with the expected theoretical result, and the noise matrix $\textbf{\textit{N}}$ identifies noise across the overall frequency modes.

As another example, we investigate a nonlinear-optical process with mode-dependent loss (Fig.~\ref{fig3}c--f)~\cite{modeloss}. Despite the loss, the amplification matrix $\textbf{\textit{A}}$ still displays the signature of a nonlinear optical process through the discrepancy between $\textbf{\textit{A}}_{ \hat{\textbf{\textit{x}}}, \hat{\textbf{\textit{x}}} }$ and $\textbf{\textit{A}}_{\hat{\textbf{\textit{p}}}, \hat{\textbf{\textit{p}}} }$. The noise matrix $\textbf{\textit{N}}$ shows non-zero off-diagonals, indicating that correlated noise is added to the output (in other words, noise is added in superposed modes). The noise matrix $\textbf{\textit{N}}$ can be further analyzed by eigendecomposition, revealing the noise eigenquadratures and the associated noise characteristics (Fig.~\ref{fig3}e,f). In Supplementary Information, we compare these results with the case of a mode-dependent loss without the nonlinear interaction. This noise analysis will help identify minimum noise channels, thereby promoting the reliable transmission of quantum information~\cite{Bachmann:2023aa,Notarnicola:2024aa}.

Finally, we characterize a nonlinear-optical process in which environmental interactions introduce quantum-correlated noise (see Fig~\ref{fig3}g,h). The characterization results show that the input field quadratures are largely attenuated while acquiring quantum-correlated noise. Regarding the noise, the process can generate entanglement into the output optical field, e.g., for the input vacuum state, the output state exhibits a negative eigenvalue of $\minus 0.37$ after partial transposition~\cite{Duan:2000fw}. This process is widely used in quantum sensing~\cite{Lemos:2014aa,wide11interferometer} and quantum communication~\cite{Madsen:2012aa,Kovalenko:2021ej,Notarnicola:2024aa} applications, as a technique to mix classical fields and quantum-correlated noise.

%\section{Discussion and Conclusion}

The ability to characterize a multimode optical process has facilitated the exploration and understanding of complex optical phenomena. Overcoming the previous limitation that focused on characterizing a linear-optical process~\cite{popoffTmat,Shapingcao,bertolotti2022imaging,mosk2012controlling,kimmaxenergy,Baek2023,Fisherinfo,Pai:2021aa,Computinggigan,quantumcomputing,universallinear,Goel:2024aa,complexlib,Integratedwang}, our work presents a general framework to completely characterize a multimode nonlinear-optical process of parametric downconversion, which finds broad applications in quantum technologies~\cite{PDC,30years,alloptical,Pereira:1994aa,Lemos:2014aa,wide11interferometer,Notarnicola:2024aa,Kovalenko:2021ej,Liu:2022eu}. Complete information has been obtained by measuring the amplification matrix and the noise matrix, revealing the complex interactions of multiple modes in a nonlinear-optical process. Furthermore, we have developed a method to decompose a multimode nonlinear-optical process, identifying eigenquadratures and the associated amplification and noise properties. This method provides a systematic way of analyzing complicated multimode nonlinear-optical processes.

The complete characterization of a linear-optical process~\cite{popoffTmat} has had far-reaching impacts on advancing photonic technologies, such as imaging through complex media~\cite{Shapingcao,bertolotti2022imaging,mosk2012controlling,Baek2023}, efficient light transmission~\cite{kimmaxenergy,Fisherinfo,Pai:2021aa}, optical computing~\cite{Computinggigan}, and quantum interference~\cite{quantumcomputing,universallinear,Goel:2024aa,complexlib,Integratedwang}. Similarly, our characterization of a nonlinear-optical process will facilitate precise control of the process, leading to unique applications beyond the reach of a linear-optical process, such as parametric amplification~\cite{alloptical} and all-optical phase conjugation~\cite{Pereira:1994aa,phaseconjugation}. The technique will play a key role in developing scalable quantum technologies, with numerous applications in quantum light generation~\cite{PDC,30years} and amplification~\cite{alloptical}, quantum nondemolition measurement~\cite{Pereira:1994aa}, quantum channel characterization~\cite{Bachmann:2023aa,Notarnicola:2024aa}, nonlinear interferometry~\cite{Lemos:2014aa,wide11interferometer}, and all-optical quantum computing~\cite{alloptical} and communication~\cite{Liu:2022eu}.

Our work focuses on the second-order optical nonlinearity leading to parametric downconversion and difference frequency generation for applications in quantum technologies. Note that the third-order optical nonlinearity of four-wave mixing involving two strong pumps leads to the same process as parametric downconversion~\cite{boyer2008entangled}, which can also be completely characterized within our framework. In addition, the current study complements recent progress in characterizing the second-order nonlinear processes of second harmonic generation and sum frequency generation for classical applications~\cite{moon2023measuring}. As a future perspective, it will be worthwhile to extend the characterization technique for higher-order optical nonlinearities~\cite{boydbook,wright2022physics}.

\section*{Acknowledgments}
This work was supported by the Ministry of Science and ICT (MSIT) of Korea (NRF-2020M3E4A1080028, NRF-2022R1A2C2006179, NRF-2023M3K5A1094806, RS-2024-00442762) under the Information Technology Research Center (ITRC) support program (IITP-2024-2020-0-01606) and Institute of Information \& Communications Technology Planning \& Evaluation (IITP) grant (No. 2022-0-01029, Atomic ensemble based quantum memory), and by the Air Force Office of Scientific Research award (FA2386-22-1-4083). MSK acknowledges support from the UK EPSRC (EP/Y004752/1, EP/W032643/1 and EP/Z53318X/1) and the National
Research Foundation of Korea (NRF) grant funded by the Korea government (MSIT) (No. RS-2024-00413957).

\setcounter{figure}{0}
\renewcommand{\figurename}{Extended Data Fig.}
\begin{figure*}
\includegraphics[width=145mm,scale=0.5]{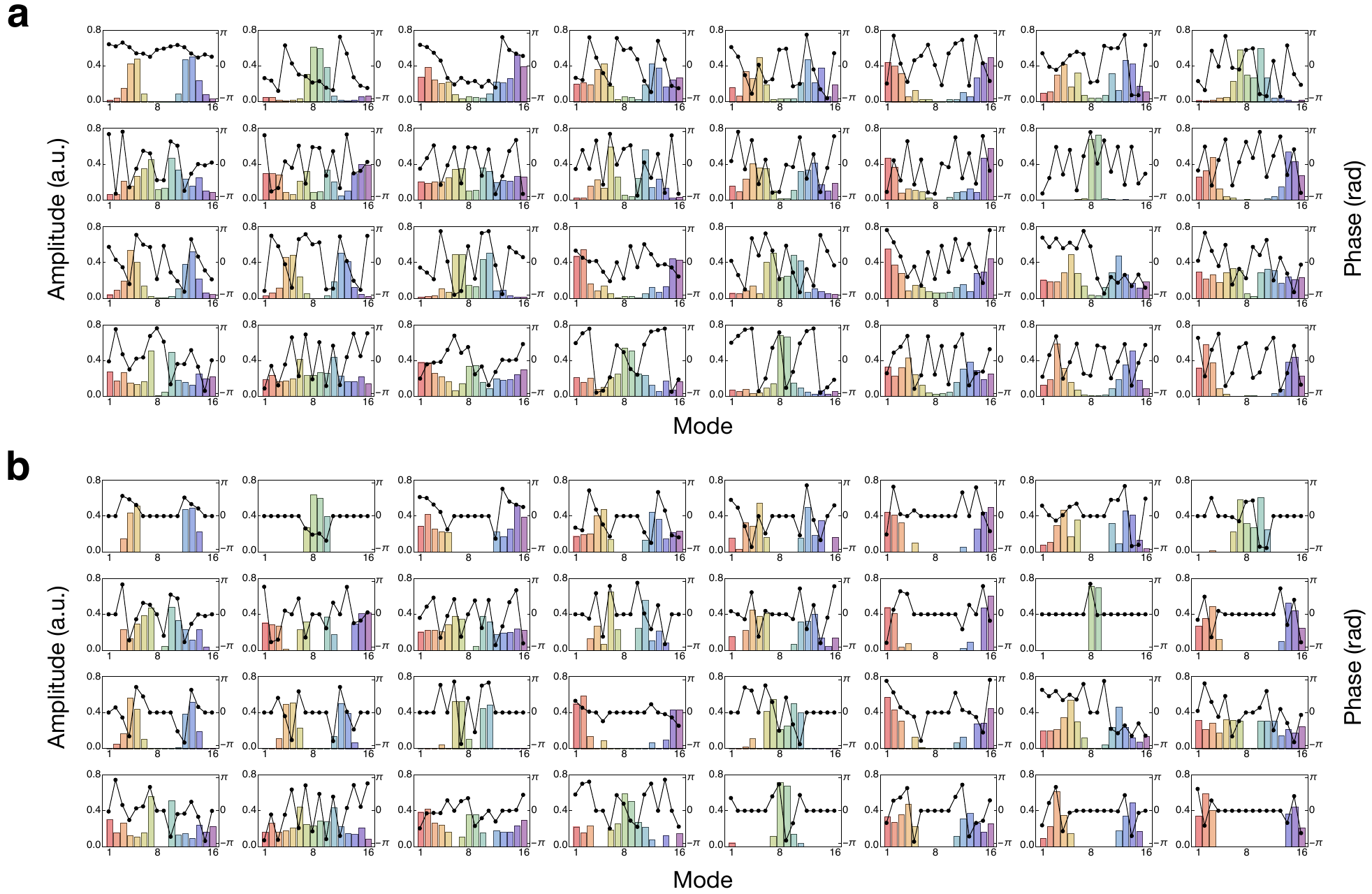}% Here is how to import EPS art
\caption{\textbf{Input eigenquadratures.} \textbf{a,} Eigenquadratures obtained by decomposing the amplification matrix (\textit{i.e.}, the column vectors of $\textbf{\textit{V}}$). \textbf{b,}  Experimentally generated eigenquadratures for probing the multimode nonlinear process. Eigenquadratures are represented by superpositions of the wavelength modes ($m=1,\dots,16$), where amplitudes and phases are obtained by $\sqrt{(q_m)^2+(q_{m+16})^2}$ and $\tan^{-1}{(q_{m+16}/{q_m}})$, respectively.
}
\label{ext_fig1}
\end{figure*}

\begin{figure*}
\includegraphics[width=145mm,scale=0.5]{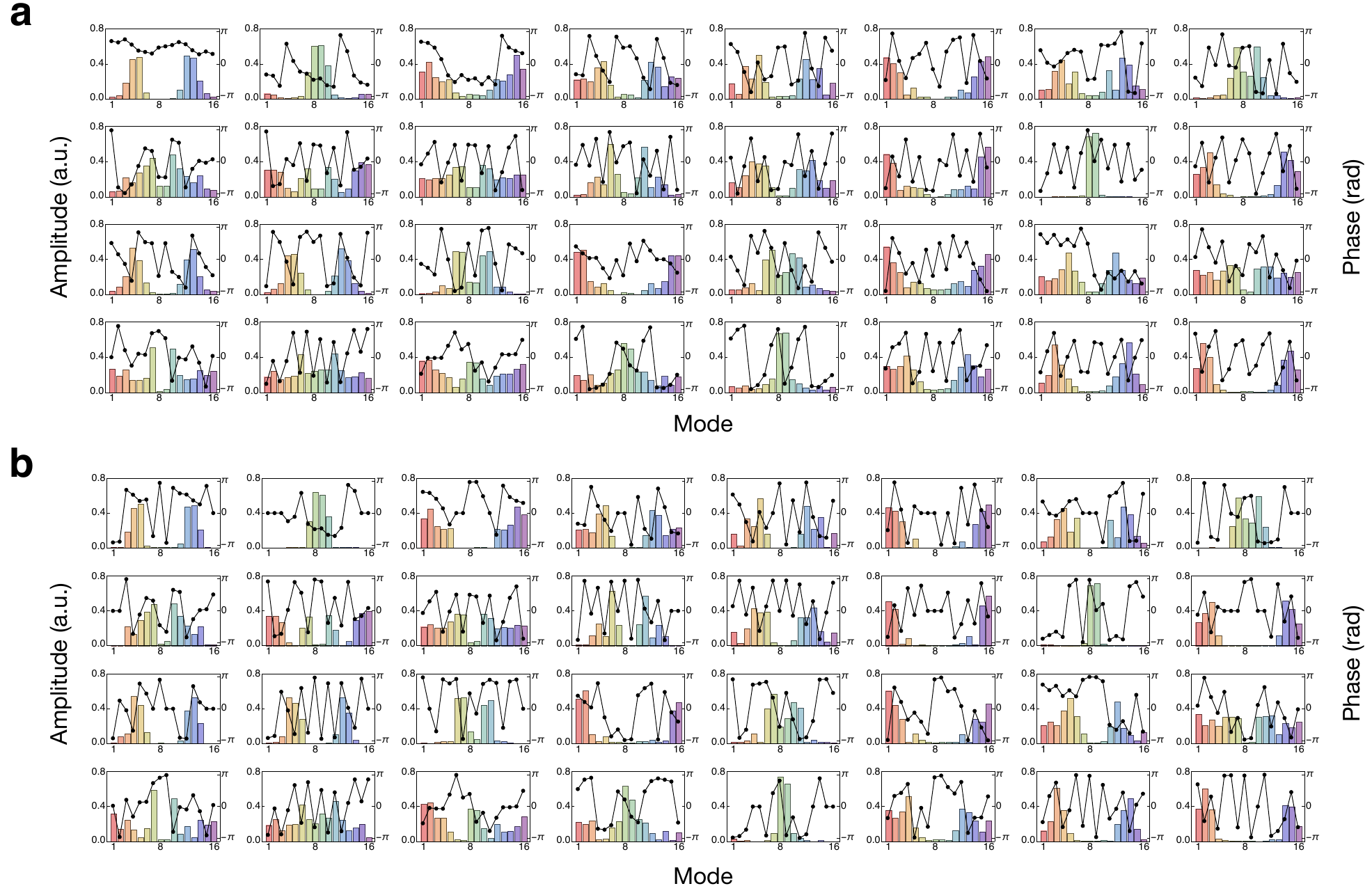}% Here is how to import EPS art
\caption{\textbf{Output eigenquadratures.} \textbf{a,} Eigenquadratures obtained by decomposing the amplification matrix (\textit{i.e.}, the column vectors of $\textbf{\textit{U}}$). \textbf{b,}  Experimentally observed eigenquadratures as a result of the multimode nonlinear process. Eigenquadratures are represented by superpositions of the wavelength modes ($m=1,\dots,16$), where amplitudes and phases are obtained by $\sqrt{(q_m)^2+(q_{m+16})^2}$ and $\tan^{-1}{(q_{m+16}/{q_m}})$, respectively.
}
\label{ext_fig2}
\end{figure*}

\end{document}